\documentclass[letterpaper,twocolumn,showpacs,preprintnumbers,amsmath,amssymb,prl]{revtex4}

\usepackage{graphics}
\usepackage{graphicx}
\usepackage{amsmath,amssymb,epsfig,picinpar,graphics,float,verbatim}

\newcommand{\dt}[1]{\frac{d#1}{dt}}

\newcommand{\ve}[1]{{#1}}
\newcommand{\nsum}[1]{\langle\,{#1}\,\rangle}
\newcommand{\ma}[1]{\mathcal{#1}}

\begin{document}

\title{Coherent regimes of globally coupled dynamical systems}
\author{Silvia De Monte$^*$, Francesco d'Ovidio$^{*,\dag}$, Erik Mosekilde$^*$}
\thanks{\{silvia,dovidio,erik.mosekilde\}@fysik.dtu.dk}
\affiliation{Chaos Group$^*$ and Center for Quantum Protein$^\dag$,
  Dept. of
Physics,\\ Technical University of Denmark, DK 2800 Kgs. Lyngby,
Denmark}
\date{\today}

\begin{abstract}
The paper presents a method by which the mean field dynamics of a population of
dynamical systems with parameter diversity and global coupling can be
described in terms of a few macroscopic degrees of freedom. The method applies
to populations of any size and functional form in the region of coherence. It
requires linear variation or a narrow distribution for the dispersed parameter.
Although being an approximation, the method allows us to quantitatively study
the collective regimes that arise as a result of diversity and coupling and to
interpret the transitions among these regimes as bifurcations of the effective
macroscopic degrees of freedom. To illustrate, the phenomenon of oscillator
death and the route to full locking are examined for chaotic
oscillators with time scale mismatch.

\end{abstract}

\maketitle

\begin{figure}[h] \center
\epsfig{file=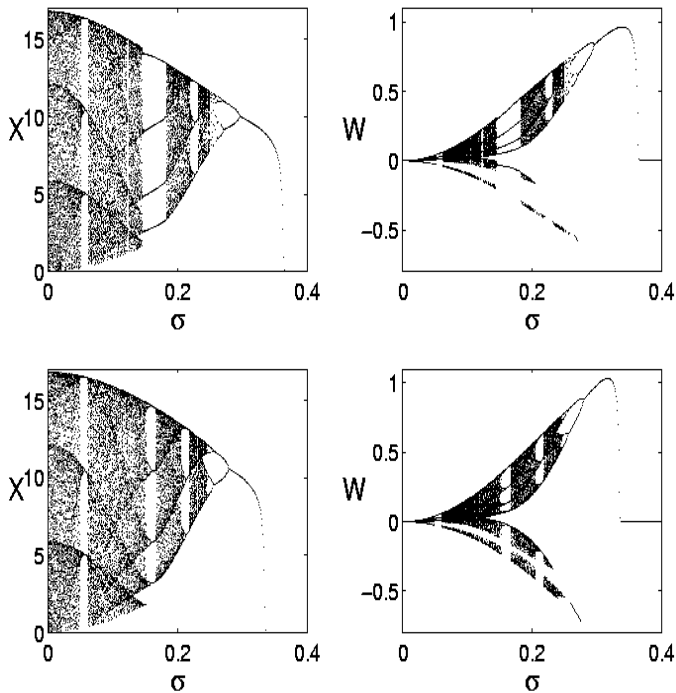,clip=,width=.5\textwidth,height=.52\textwidth}
\caption{Poincar\'e section of the mean field and of the shape
 parameter for a population of 32 coupled
R\"ossler oscillators ($a=0.25$, $b=1$, $c=8.5$) with time scale
mismatch (top) and for its order parameter reduction (bottom). The
 coupling is $k=1$. 
The systems go from chaos to oscillator death when the standard
 deviation $\sigma$ of the parameter distribution increases. The
 reduced system reproduces the bifurcation cascade of the population
 with remarkable quantitative agreement.
\label{fig:feig} }\end{figure}

{\it Introduction.} Populations of globally coupled dynamical systems
represent a 
useful framework to study the collective properties of biological
systems \cite{biolsyst}, allosterically activated enzymatic
reactions \cite{stange98}, electronic devices
\cite{hadley88}, and chemical reactions
\cite{chemreact,bar-eli85,kuramotobook}. Although started as
qualitative analyses, in the last years it has appeared that such
theoretical approaches can be linked quantitatively to experimental
systems, like arrays of Josephson junctions \cite{strogatzjj} or metabolic
synchronization in suspensions of yeast cells \cite{yeast}. Globally
coupled systems can show different macroscopic behavior when the coupling
strength 
and the parameter dispersion are changed. 
If the coupling is high
enough and the parameter dispersion sufficiently small, the elements of the
population evolve in time close to each other (and thus to the mean field) in
phase space. In the opposite case, the elements of the population move
incoherently and eventually their positions average out, so that the 
asymptotic dynamics of the mean field is characterized only by fluctuations 
that vanish in the limit of an infinite number of elements. 
Between these two limiting cases, complex collective behavior
arise. This type of scenario has been observed in
a wide number of systems, starting from the early works of Winfree and Kuramoto
on phase rotators \cite{rotators}, over limit
cycle oscillators \cite{strogatz00} to chaotic oscillators
\cite{chaotic}. Investigations have been performed with various
statistical methods, based on   
phase reduction \cite{kuramotobook,daido97}, the continuity equation formalism 
\cite{matthews90}, or the slaving principle \cite{shimizu84}.
In particular, the introduction of order parameters has appeared
to be useful for quantifying the collective regimes, as it aims at providing
a direct link between the microscopic and the macroscopic dynamics. 
However, due to the fact that for low coupling the system spans a region 
of phase space whose dimension grows with the population size, the onset
of macroscopic oscillations from \emph{incoherence} cannot, in general, be 
described by means of a closed system of a few macroscopic degrees of freedom.
In the \emph{coherent} region, on the other hand, the
macroscopic dynamics lives on a low dimensional manifold which
coincides, in the limit of identical oscillators, with the trajectory
of a single uncoupled element of the population, and thus a description
in terms of a few effective degrees of freedom may work. 
Indeed, for limit cycle oscillators with strong coupling and small 
parameter diversity, an order parameter expansion was recently given by 
De Monte and d'Ovidio \cite{demonte02}, showing that the transient and 
asymptotic dynamics of the mean field can be accounted for by only two 
macroscopic variables.
In the present work we propose a method by which the mean field
dynamics of globally coupled dynamical 
systems (maps or ODEs) with dispersion on one parameter can be
systematically reduced (in the coherent regimes) to a system of two coupled
order parameters.
The conditions we need to impose are either a linear
dependence or a narrow distribution for the dispersed parameter and, in
general,
smooth dynamical systems. The method can be applied to both finite and infinite
population sizes.
The paper is organized in two parts. First, we
generalize the method proposed in Ref. \cite{demonte02} to any population of
ODEs or maps:
the equations of motion of the mean field and of a second order parameter,
appearing when the parameter diversity is introduced, are derived from
the microscopic equations through a perturbative approach. 
The theory provides approximate but still quantitative predictions,
which are tested on populations
of ODEs with time scale mismatch. In particular, we study the phenomenon of 
\emph{oscillator death}, in which an attracting equilibrium appears in the 
system due to the interplay between high parameter dispersion and strong coupling. 
We then show that the approach continues to apply to non-stationary
regimes: varying the parameter spread, the transition from oscillator death to
full synchronization in a population of R\"ossler systems is followed,
and this allows us to
characterize the various locked regimes in terms of a cascade of macroscopic
bifurcations.\\
{\it Order parameter expansion.} Let us consider a population of $N$
dynamical systems in the coherence region, i.e., such that the
distance $||X-x_j||$ between the mean field $X$ and the state $x_j$ of
any of the oscillators remains small in time. 
For ODEs the dynamics of the j-th element of the population is defined by the
equation: \[ \dt{x_j}=f(x_j,p_j)+K(X), \] while for maps: \[
x_j\to f(x_j,p_j)+K(X). \] To simplify the notation, we will write
in general $\ma{T}{x_j}=f(x_j,p_j)+K(X)$ where $\ma{T}x_j$ is the
operator of time evolution applied to $x_j$.
The variables
$x_j\in \mathbb{R}^n$ are the state vectors and $p_j\in\mathbb{R}$ is a
real parameter taken from a distribution of average $p_0=\nsum{p}$
and variance $\sigma^2=\nsum{(p-p_0)^2}$.
The oscillators are assumed to be globally coupled to the
mean field $X=\nsum{x}\equiv \sum_{j=1}^N x_j/N$
through the function $K:\mathbb{R}^n\rightarrow \mathbb{R}^n$. A
change of variables $x_j=X+\epsilon_j$ expresses the position of every element
of the population in terms of the mean field, so that in the coherent
regime a series expansion in 
$\epsilon_j$ can be performed. Setting $\delta_j=p_j-p_0$ and writing the mixed
differentials in $x$ and $p$ as $\ma{D}_{x,p}$, we get:
\begin{eqnarray}\label{eq:onexp}
\ma{T}x_j=&f(X,p_0)+\ma{D}_{x}f(X,p_0)\epsilon_j+
\ma{D}_{p}f(X,p_0)\delta_j\nonumber\\&+
\ma{D}_{x,p}f(X,p_0)\epsilon_j\delta_j+o(|\epsilon_j|^2)+C.
\end{eqnarray}
The terms in $o(|\epsilon_j|^2)$ are small because of the hypothesis of
coherence. $C$ contains all the Taylor terms that are not
$O(|\epsilon_j|^2)$, i.e., terms with a factor of the type:
$\ma{D}_{x,p_m}\delta^m$ with $m>1$. This term may be neglected if
the dependence on the parameter is linear
or if the parameter distribution is narrow. In fact, in the first
case $\ma{D}_{x,p_m}\delta^m=0, \quad \forall m>1$, while in the second the
term
is $O(\delta_j^2)$. We can thus discard these two terms and construct the
macroscopic equations, by averaging Eq.\ (\ref{eq:onexp}) and using
$\nsum{\epsilon}=\nsum{\delta}=0$:
\begin{eqnarray*}
\ma{T}{X}=f(X,p_0)
+\ma{D}_{x,p}f(X,p_0)\nsum{\delta\,\epsilon}. \end{eqnarray*}
The only first order term that is left defines a new macroscopic variable
$W:=\nsum{\delta\,\epsilon}$, that we will call the \emph{shape
  parameter}. It measures the dispersion in both
the parameter and the phase space.\\
In order to close the system, the equation of motion for
$W$ has to be obtained as well. As for the mean field, this can be done
by writing:
$\ma{T}W=\ma{T}\nsum{\delta\,\epsilon}=\nsum{\delta \ma{T}x}$ and then using
Eq.\
(\ref{eq:onexp}). Discarding again higher order terms and making the
closure assumption that the term $\nsum{\epsilon\delta^2}$ is negligible
relative to $\nsum{\delta^2}$ (which is true in the coherent regimes), 
a closed (approximated) system is obtained for the two macroscopic
variables:
\begin{eqnarray}\label{eq:odes} \begin{cases}
\ma{T}{X}=f(X,p_0)+\ma{D}_{x,p}f(X,p_0)W+K(X)\\
\\ \ma{T}{W}=\sigma^2\,\ma{D}_{p}\,f(X,p_0)
+\left[\ma{D}_{x}\,f(X,p_0)\right]W. \end{cases}
\end{eqnarray}
The mean field thus behaves like an uncoupled individual element if the
oscillators are identical (since in this case $W=0$). However,
parameter diversity may induce new regimes.
We remark that the method can be applied in the same way for other
choices of the coupling term.\\
{\it Populations with time scale diversity.}
Let us now apply the order parameter reduction to
the coherent behavior in populations of oscillators with time scale
diversity. These oscillators may be regarded as a generalization of
oscillators with 
different natural frequencies (e.g., the Kuramoto model) and are
defined by the equations:
\begin{equation}
\label{eq:timesc}
\dt{\ve{x}_j}=\tau_j\,\ve{g}(\ve{x}_j)+\ma{K}\,(\ve{X}-\ve{x}_j).
\end{equation}
$\tau_j$ are strictly positive numbers (which rescale the speed along
the orbit) and the matrix $\ma{K}=k\ma{I}$
provides an isotropic diffusive coupling. \\
{\it Oscillator death.} 
As a first application, we will address the phenomenon
of oscillator death, that was first described in limit cycle
oscillators with a
large natural frequency mismatch and strong coupling
\cite{oscdeath,shimizu84}. Under 
these conditions, all the elements eventually collapse on the origin, which 
is an unstable focus for the uncoupled systems. It was then supposed that the 
phenomenon is more general \cite{Ermentrout90} and it was 
actually described in many other systems (like Brusselators \cite{bar-eli85}
and biological systems \cite{sherman98}) where an equilibrium, unstable
in the uncoupled case, becomes attracting and suppresses the periodic
or chaotic oscillations. 
When the oscillator death takes place, the regime is trivially 
coherent since all the elements of the population lie on the same
equilibrium (which does not need to be unique) and the dependence on the
dispersed parameter is 
linear, so that the requirements for applying the order parameter
reduction are both satisfied.
The order parameter expansion will allow us to treat the problem in
general and with a  
simple analysis, showing that an unstable equilibrium of $g$ (that,
without loss of generality, we put at the origin ) can 
always be stabilized by large time scale dispersion and high coupling, 
provided that the equilibrium is a saddle-focus whose repelling eigenvalues
have an imaginary part larger than the real one. 
In the reduced system Eq.\ (\ref{eq:odes}), the macroscopic equilibrium
corresponding to oscillator death is the point
$X=0,\, W=0$ and its stability can be studied with simple algebra (the Jacobian
matrix is composed by four blocks which can be simultaneously
diagonalized). For each eigenvalue $\lambda^l=a_l+\,i b_l$ of the
unstable equilibrium of $g(x)$, two macroscopic eigenvalues appear:
\begin{eqnarray}\label{eq:eigmacr}
\Lambda^{l}_{1,2}=\lambda_l-\frac{k}{2}\pm\,\sqrt{\left(\frac{k}{2}\right)^2+
\lambda_l^2\,\sigma^2}
\end{eqnarray}
It is clear that, if the equilibrium for the uncoupled system is repelling due
to a purely real and positive eigenvalue, then the corresponding
$\Lambda$ always 
has a positive real part and the oscillator death solution cannot be stable.
Instead, if the instability is due to couples of complex conjugates
the condition for the oscillator death to be stable is that the real
parts of the eigenvalues (\ref{eq:eigmacr}) are all negative.\\
Let us now consider the case in which
only two eigenvalues of the uncoupled system are complex conjugate, while the
others are real and negative. This is the only case where oscillator
death can appear in 
populations of globally coupled two- or three-dimensional systems, like coupled
Lorenz or R\"ossler oscillators.
\begin{figure}[h] \center
	\epsfig{file=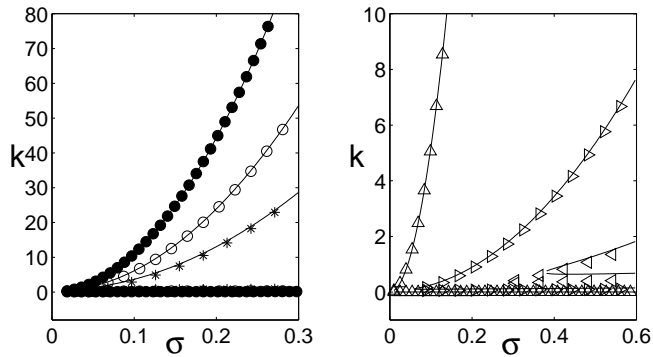,clip=,width=.48\textwidth}
	\caption{Comparison of the bifurcation boundaries for the oscillator death in
the full and the reduced system
	 for different populations of dynamical systems: Lorenz
	 ($p=10$, $b=8/3$) $r=28$ ($\bullet$), $r=32$ ($\circ$),
	 $r=50$ ($\ast$); 
	 R\"ossler ($b=.4$, $c=8$) $a=0.01$ ($\triangle$), $a=0.1$
	 ($\triangleright$), $a=0.4$ ($\triangleleft$). 
	 The continuous lines are the bifurcation boundary for the
	 corresponding reduced systems. Populations of 100 elements and
	 Gaussian parameter distributions have been used.}
	\label{fig:adregion}
\end{figure}
The bifurcation condition Eq.\ 
(\ref{eq:eigmacr}) now reduces to a single inequality.
Recalling the definition of $\lambda$, the
boundary of the oscillator death $\text{Re}\left(\Lambda\right)=0$ (where the indexes
have been dropped due to the fact 
that there is only one pair of complex eigenvalues) can be rewritten
in the form:  
\begin{eqnarray}\label{eq:bifcond}
\left|\:\text{Re} \left( \sqrt{1+4\left(1+i\gamma\right)^2\,\left(\frac{\sigma}{k^*}
\right)^2}\right)\right|
=1-\frac{2}{k^*}.
\end{eqnarray}
Here $k^*=k/a$ and $\gamma=b/a$, corresponding to a rescaling of the variables
so
that the real part of the uncoupled system's eigenvalues is unitary.
\begin{figure}[h] \center
\epsfig{file=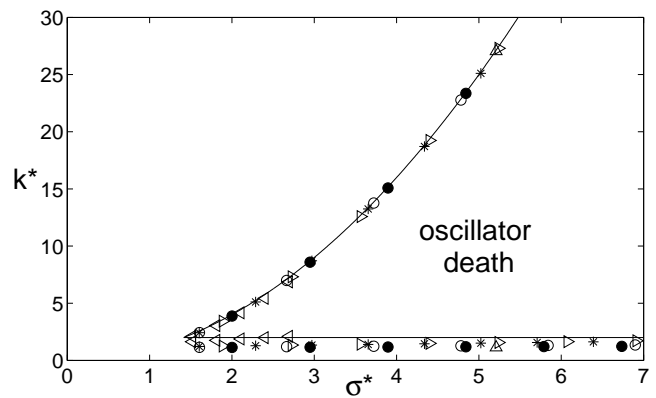,clip=,width=.48\textwidth}
\caption{The bifurcation boundaries of oscillator death for different types of
oscillators collapse on the same, analytically determined curve
(continuous line) after rescaling the parameters $k$ and $\sigma$. Symbols as
in
Fig.\ \ref{fig:adregion}. 
\label{fig:univ}} \end{figure}
The bifurcation boundary is thus a surface in the space ($k^*$,
$\sigma$, $\gamma$). The fact that only three parameters account for the
stability of oscillator death in any population with the aforementioned
characteristics follows from the fact that oscillator death is essentially
a local phenomenon, so that the nonlinearities influence at most the critical
character of the macroscopic Hopf bifurcation.
In the plane ($k^*$,
$\sigma$), the bifurcation boundaries are lines parameterized by
$\gamma$ (Fig.\ \ref{fig:adregion}).
Their asymptotic behavior can be computed from Eq.\ (\ref{eq:bifcond})
and gives $k^*=(\gamma^2-1)\,\sigma^2$ (upper boundary, to locking) and
$k^*=2(1+\sigma$) (lower boundary, to incoherence).
Moreover, if the imaginary part
of $\lambda$ is much larger than the real one (i.e., in the limit
$\gamma\to\infty$) a further rescaling is
possible: $\sigma^*=\sqrt{\gamma^2-1}\,\sigma$. In this limit, the
bifurcation boundaries can be rescaled
to the same line: $k^*=\sigma^{*2}$ and $k^*=2$.
Oscillators as different
as R\"ossler and Lorenz systems rescale to the same curve
obtained analytically from Eq.\ (\ref{eq:bifcond}) (Fig.\ \ref{fig:univ}).\\
{\it Non-steady coherent regimes.} Let us now speculate about what
happens outside the 
region of oscillator death. On the lower boundary the coupling is weak
relative to the parameter spread and the system will bifurcate into
incoherence, therefore we expect the expansion to no longer be valid. On the
upper boundary, by contrast, where the closure assumptions are satisfied, the
phase transition to the locked state arises macroscopically as a Hopf
bifurcation. The validity of the introduced approximations, moreover,
is mantained in the locked regime, where the oscillators remain close to
the mean field. For this
reason, not only the transition out of the oscillator death, but \emph{any
other bifurcation up to the fully synchronized state can be identified
through the reduced system}.
As an example, Fig.\ \ref{fig:feig} compares the behavior of the mean field
of a population of 32 R\"ossler oscillators with time scale
mismatch to the order parameter expansion Eq.\
(\ref{eq:odes}). Changing the spread $\sigma$ in the time scale
distribution, a complete period doubling cascade can be followed, connecting 
the fully locked regime to oscillator death, with
remarkable quantitative agreement.\\ 
From the validity of the order parameter expansion in describing the
macroscopic chaotic dynamics one can infer that the Lyapunov spectrum
of the reduced system is composed by the dominant eigenvalues of the
full system. The microscopic degrees of freedom which are neglected in
the approximation would therefore act as a perturbation on the
collective dynamics and give a small contribution to the
Kaplan-Yorke dimension of the macroscopic attractor.\\    
{\it Conclusions.} In ensembles of globally coupled oscillators, 
parameter diversity may induce nontrivial collective
behavior, where the mean field dynamics is qualitatively different
from that of each uncoupled element.
In this work we have shown that such regimes are low dimensional in the region
of coherence, and that a description by means of few effective degrees of
freedom may be given. This is done through an expansion around the perfectly
synchronized state. Although the method involves some approximations (it is
not exact even for a linear analysis), it provides nevertheless an accurate
and quantitative description of the dynamics at the macroscopic level. We
remark that there are no requirements on the population size: as far as the
oscillators are of the same type, populations of different sizes behave the
same if they have the same coupling term and the
same variance of the parameter distribution. The somewhat surprising
consequence of this fact is that the macroscopic features of the
coherent regimes can
be accounted for by a system of just \emph{two} coupled oscillators,
providing a two-body
approximation of the population dynamics. 
Finite size effects, however, do arise as the region of
incoherence is approached, and the dimension of the collective
dynamics increases. 
There are several ways in which we think our
approach can be
developed further. In particular, the inclusion of higher order terms may
allow us to explain more complex collective regimes arising close to
incoherence.
Moreover, a similar approach accounts for the
effect of noise on the collective dynamics of identical oscillators.
A macroscopic
bifurcation scenario similar to that induced by parameter diversity
appears in large populations of noisy chaotic maps. This bifurcations
and the finite size effects can again be described in the 
framework of an order parameter expansion \cite{demonte02u}. \\
S.\ D.\ M. and F.\ d'O. are thankful to the organizers of the NATO School
on Synchronization: Theory and Application (May 19-June 1 2002, Yalta,
Ukraine), where these results found their final shape. F. d'O. acknowledges 
support from the Danish National Research Foundation.

\bibliography{bibod}

\end{document}